\documentclass[a4paper, 11pt, oneside]{memoir}

\usepackage[utf8]{inputenc}
\usepackage[english]{babel}
\usepackage[T1]{fontenc}

\setsecnumdepth{subsection}
\settocdepth{subsection}
\chapterstyle{hangnum}
\usepackage{microtype}

\usepackage{relsize}

\usepackage[bottom,perpage,symbol]{footmisc}

\checkandfixthelayout[nearest]

\usepackage{caption}

\usepackage[margin=10pt, font=small, labelfont=bf, labelsep=endash]{caption}

\usepackage{dirtree}

\usepackage{color}
\usepackage{xcolor}

\newcommand{\filehdf}[1]{\textcolor[rgb]{0.55, 0.55, 0.55}{{#1}}}
\newcommand{\filec}[1]{\textcolor[rgb]{0.99, 0, 0.0}{{#1}}}
\newcommand{\filedir}[1]{\textcolor[rgb]{0.1, 0.4, 0.95}{{#1}}}
\newcommand{\filepy}[1]{\textcolor[rgb]{0.9, 0.3, 0}{{#1}}}
\newcommand{\filemakefile}[1]{\textcolor[rgb]{0.65, 0, 0.7}{{#1}}}
\newcommand{\filebash}[1]{\textcolor[rgb]{0.2, 0.6, 0}{{#1}}}
\newcommand{\filetext}[1]{\textcolor[rgb]{0, 0, 0}{{#1}}}
\newcommand{\filepdf}[1]{\textcolor[rgb]{0.75, 0.75, 0}{{#1}}}

\usepackage[colorlinks=true, linkcolor=black, citecolor=black, urlcolor=blue]{hyperref}

\usepackage[autostyle]{csquotes}
\renewcommand{\quote}[1]{\foreignquote{english}{#1}} 

\usepackage[xspace]{ellipsis}

\usepackage{listings}

\usepackage{enumitem}

\usepackage{slantsc}

\usepackage{amsmath}

\usepackage{amssymb}
\usepackage{mathtools}

\mathtoolsset{showonlyrefs}

\usepackage[load-configurations=abbreviations,separate-uncertainty=true,binary-units=true]{siunitx}

\usepackage{mathdots}

\usepackage{nicefrac}

\usepackage{isomath}

\allowdisplaybreaks

\makeatletter
\newcommand{\partsmash}[2][tb]{%
  \def\mb@t{\ht\z@ #2\ht\z@}\def\mb@b{\dp\z@ #2\dp\z@}%
  \def\mb@tb{\mb@t \mb@b}%
  \edef\finsm@sh{\csname mb@#1\endcsname\box\z@}%
  \ifmmode \@xp\mathpalette\@xp\mathsm@sh
  \else \@xp\makesm@sh
  \fi}
\makeatother

\usepackage{xspace}
\newcommand{\concept}{\textsc{co\textsl{n}cept}\xspace}
\newcommand{\Concept}{\textsc{Co\textsl{n}cept}\xspace}

\usepackage[pdftex]{graphicx}

\begin{document}
\date{October 19, 2015}
\title{\Huge\textbf\concept \\[0.5em] \LARGE The COsmological $N$-body CodE in PyThon \\[1em] \Large\textbf{User's Guide}}
\author{\large Jeppe Mosgaard Dakin\\[0.6em] \normalsize jeppe.mosgaard.dakin(at)post.au.dk}
\maketitle

\begin{center}
\vspace{-1.5em}
\includegraphics[width=0.75\textwidth]{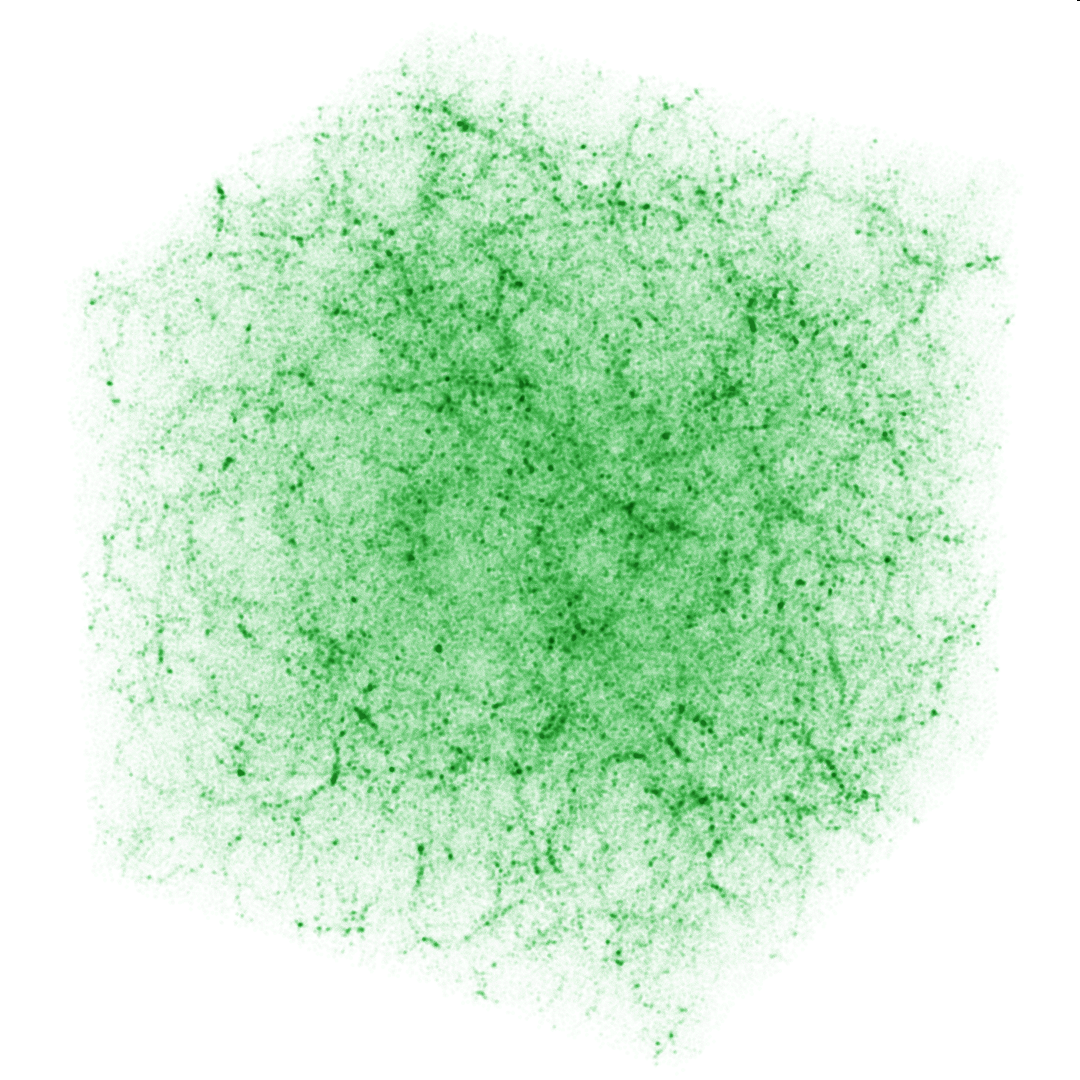}
\vspace{-0.5em}
\end{center}

\begin{center}
This document serves as a user's guide to the \concept code. It is split up into two parts, the first of which gives a basic overview of the functionality of the program, while the second is much more in-depth.

The GitHub page for the \concept code can be found at \\ \url{https://github.com/jmd-dk/concept/}
\end{center}

\newpage
\tableofcontents*

\part{Overview}
\chapter{Introduction}
\Concept (COsmological $N$-body CodE in PyThon) is a free and open-source code for cosmological $N$-body simulations on massively parallel computers with distributed memory. As of the time of writing, collisionless dark matter is the only implemented particle species. Gravity can be computed using the PP, PM or the $\text{P}^{3}\text{M}$ algorithm. The goal of \concept is to make it pleasant to work with cosmological $N$-body simulations --- for the cosmologist as well as for the source code developer.

\Concept is easy to use and has good performance. The source code is properly structured, well-commented and consist almost exclusively of \textsc{pep} {\smaller 8} compliant\footnote{\textsc{Pep} {\smaller\smaller 8} is the official style guide for writing clean and idiomatic Python code.} Python code.

\Concept relies on the Cython compiler, which translates (or \emph{cythonizes}) Python source code to equivalent C source code, which are then compiled as any other C program. The computationally expensive parts of the code are written such that they translate very directly into C code, while the inexpensive parts are written using the whole spectrum of modern Python constructs, which are only poorly translated into C code\footnote{When no equivalent C construct exists for a given Python construct, Cython simply generates the CPython C \textsc{api} calls corresponding to the given Python construct.}. In effect this allow for high performance \emph{and} rapid development.

In addition to running \concept in compiled mode, it is also possible to run it as any other Python script, though with a hefty performance penalty.

The \concept code is written in version 3 of the Python language. Among other things, this allows for Unicode literals in strings and variable names, which \concept takes advantage of. All source files are therefore encoded in \textsc{utf-}{\smaller 8}. This unconventional and slightly dangerous choice is made in the name of æsthetics.

Although \concept is intended for computer clusters, it really runs on anything down to single-core laptops. Communication between \textsc{cpu}s is done explicitly by means of the message passing interface (\textsc{mpi}). The code has been tested and found to work on a large number of Linux systems.

\chapter{Basic Usage}
Here we state the very basic functionalities of \concept in a tutorial-like manner\footnote{If this is your first experience with \concept, it is advised that you try it out on a local computer rather than on a remote host, as the behavior of \concept differs somewhat for these two cases.}. For a much more in-depth description, see part \ref{part:indepth} of this document.

\section{Installation}
To install \concept along with all the dependency programs it needs, simply run the command
\begin{lstlisting}[language=bash, backgroundcolor=\color{yellow!20}, xleftmargin=\parindent, xrightmargin=\parindent, basicstyle=\small\ttfamily, columns=fullflexible]
bash <(wget -O- --no-ch bit.ly/concept_nbody)
\end{lstlisting}
from a terminal, which will download and launch the \concept installer. You will be prompted for an installation directory in which everything will be placed. To uninstall \concept, simply delete this directory. The installation will consume slightly less than \SI{1}{\giga\byte} of disk space and will complete within \SIrange{1}{2}{hours} on modern hardware.

\section{Files and Directories}
\begin{figure}[tb]
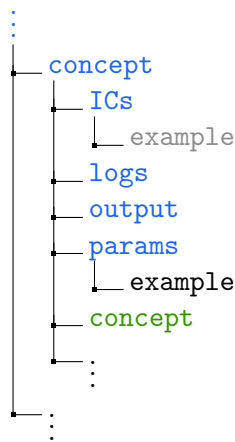

\center
\begin{minipage}{0.3\linewidth}
\dirtree{%
.1 \hspace{3pt}\smash{\filedir{$\vdots$}}.
.2 \filedir{concept}.
.3 \filedir{ICs}.
.4 \filehdf{example}.
.3 \filedir{logs}.
.3 \filedir{output}.
.3 \filedir{params}.
.4 \filetext{example}.
.3 \filebash{concept}.
.3 \raisebox{-0.84em}{\filetext{$\vdots$}}.
.2 \raisebox{-0.84em}{\filetext{$\vdots$}}.
}
\end{minipage}
\caption{Basic directory tree for \concept. The colors indicate directories (blue), Bash scripts (green), text files (black) and \textsc{hdf}{\smaller 5} files (gray).}
\label{fig:basicdirtree}
\end{figure}

A stripped down version of {\concept}'s directory tree is shown in figure \ref{fig:basicdirtree}. The \texttt{concept} directory contains four subdirectories and the \texttt{concept} executable, which is just a Bash script. The \texttt{ICs} subdirectory is intended for storage of IC files. As part of the \concept distribution, an example IC file is provided. This IC file is written in {\concept}'s so-called \quote{standard} \textsc{hdf}{\smaller\smaller 5} format. Similarly to \texttt{ICs}, the \texttt{params} subdirectory is intended for the storage of parameter files. An example parameter file, which is just a text file of definitions, is provided. Parameter and IC files are described in detail in chapters \ref{ch:params} and \ref{ch:snapshots}, respectively.

The \texttt{output} directory is the default directory for \concept to dump its output, while \texttt{logs} is where it save log files. Everything printed to the screen in each run will be logged automatically.

\section{Running the Code}\label{sec:running_partI}
The simplest possible invocation of {\concept} --- assuming that you are in the \texttt{concept} directory --- is
\begin{lstlisting}[language=bash, backgroundcolor=\color{yellow!20}, xleftmargin=\parindent, xrightmargin=\parindent, basicstyle=\small\ttfamily, columns=fullflexible]
./concept
\end{lstlisting}
You will be told that you did not specify neither the number of processes nor the parameter file to use. \Concept then resorts to running a single process and using its internal default values for all parameters. If this is the first run or the code is not built for some other reason, \concept will now do so. When the build process is completed, the actual run will begin and immediately finish, without really doing anything.

To control the number of processes to use and specify a parameter file, invoke \concept like e.g.\
\begin{lstlisting}[language=bash, backgroundcolor=\color{yellow!20}, xleftmargin=\parindent, xrightmargin=\parindent, basicstyle=\small\ttfamily, columns=fullflexible]
./concept -n 2 -p params/example
\end{lstlisting}
which will use two processes to run the simulation defined by the example parameter file. As ordered by the parameter file, a snapshot will be produced in the \texttt{output} directory.

The \texttt{concept} script has many more options, which will be described in detail in section \ref{sec:commandlineinterface}. For now you may take a quick look at the available options using
\begin{lstlisting}[language=bash, backgroundcolor=\color{yellow!20}, xleftmargin=\parindent, xrightmargin=\parindent, basicstyle=\small\ttfamily, columns=fullflexible]
./concept -h
\end{lstlisting}

\subsection{Parameters and Output}
Let us take a short look at some of the parameters in the example parameter file. Opening the file we see that the parameters are grouped into categories. In the first category, input/output, the IC file together with any wanted output is specified. Below some of the parameters in this category are given new values:
\begin{lstlisting}[language=bash, xleftmargin=\parindent, xrightmargin=\parindent, basicstyle=\small\ttfamily, columns=fullflexible, frame=single, framexrightmargin=0em, columns=fullflexible, keepspaces=true, mathescape]
# Input/output
IC_file = 'ICs/example'
snapshot_type = 'standard'
output_dirs = {
    'snapshot' : 'output',
    'powerspec': 'output/powerspec',
    'render'   : '',
               }
output_bases = {
    'snapshot' : 'spam',
    'powerspec': 'ham',
    'render'   : 'eggs',
                }
output_times = {
    'snapshot'       : 1,
    'powerspec'      : (0.5, 1),
    'render'         : linspace(0.05, 1, 5),
    'terminal render': [sqrt(i) for i in (0.3, 0.4, 0.5)],
                }
\end{lstlisting}
As listed in the \texttt{output\_times} variable, four different kinds of output are available:
\begin{description}
    \item[Snapshots] \hfill \\
    This is simply a dump of the raw state of the particle system, in the format specified by \texttt{snapshot\_type}.
    \item[Power spectra]  \hfill \\
    Instead of simply outputting the raw data, you may be more interested in some reduced form of it. The power spectrum is one such form. The primary power spectrum output is a text file containing the power spectrum data. Optionally, \concept is able to also plot the power spectrum and produce an image file (look for the \texttt{powerspec\_plot} parameter in the graphics category). Throughout the \concept code, a power spectrum is referred to as a \emph{powerspec}.
    \item[Renders] \hfill \\
    \Concept is able to produce 3D graphical visualizations of the particles. Each render output is an image of one such visualization. The appearance of the renders can be adjusted using parameters in the graphics category. 
    \item[Terminal renders] \hfill \\
    Like renders, terminal renders are graphical visualizations of the particles. Unlike renders however, terminal renders get printed directly in the terminal and are not saved to disk (disregarding the log file). The appearance of the terminal renders can be adjusted using parameters in the graphics category. 
\end{description}
Since all four kinds of output have non-empty output times in the parameter file snippet above, they are all going to be produced. An output will be produced during the simulation when the scale factor matches a value given in \texttt{output\_times}. As demonstrated above, \concept accepts values for \texttt{output\_times} in a variety of different formats. It holds generally for all parameters that any meaningful and valid Python statement may be used as their definition.

The values in \texttt{output\_dirs} are the directories in which to dump the different kinds of output. These --- as well as all other relative paths in the \concept code --- should be stated relative to the \texttt{concept} directory. In the parameter file snippet above, we see that snapshots will be placed in the standard \texttt{output} directory, while power spectra will be placed in the non-standard \texttt{output/powerspec} directory. Any non-existing output directories will be created as needed. Lastly, we see that no output directory is assigned to renders. As renders are going to be produced, \concept assigns \texttt{output} as the render output directory. This is an example of {\concept}'s use of default parameter values. You can leave out any parameter from your parameter files and they will be handled for you in a similar manner. It is however not generally advised to write parameter files which depend on the choice of such default assignments.

The values in \texttt{output\_bases} serves as filename prefixes given to the different output. This together with the output time labels the output files. The parameter file snippet above will e.g.\ produce the output files \texttt{output/spam\_a=1.00}, \texttt{output/powerspec/spam\_a=1.00} and \texttt{output/eggs\_a=1.00.png}.

\part{In-depth Guide}\label{part:indepth}
\chapter{Installation}
The \concept code should be able to compile and run on virtually any Linux system. The software stack needed to build and run the code is quite large, taking up nearly \SI{1}{\giga\byte} of disk space. To ease the process of installation of this stack, \concept comes with an \texttt{installer} script, which automates this process.

Some basic system dependencies\footnote{By \quote{system dependencies} we mean software components which are usually installed system-wide by the root user. The complete list of system dependencies are \textsc{awk}, Bash (version 4.0 or newer), sed, \textsc{gnu} Coreutils, \textsc{gnu} Make, \textsc{gcc}, GFortran, G++, gzip, Perl, tar and wget. Only the first six of these are strictly necessary to run the \textsc{co\textsl{n}cept} code, though all of them are needed to run the \texttt{installer} script.} are required to run even the \texttt{installer}. These programs are pre-installed on most Linux systems. If \texttt{installer} is run while some system dependencies are missing, these will be reported to you. Additionally, if the \texttt{installer} finds an installed package-manager, it will prompt for installation of the missing system dependencies (requiring root privileges).

\section{Dependencies}\label{sec:dependencies}
The \textsc{co\textsl{n}cept} code has the following free and open-source software dependencies:
\begin{description}
    \item[\textbf{\textsc{gsl}}] The \textsc{gnu} Scientific Library.
    \item[\textbf{\textsc{mpi}}] Any implementation of the Message Passing Interface (version 3.0 or newer).
    \item[\textbf{\textsc{hdf}{\smaller\smaller 5}}] The Hierarchical Data Format, linked to the \textsc{mpi} library and configured to be parallel. It has the zlib library as a dependency.
    \item[\textbf{\textsc{fftw}}] The Fastest Fourier Transform in the West (version 3.3 or newer), linked to the \textsc{mpi} library and configured to be parallel.
    \item[\textbf{Python}] The standard C-implementation of the Python programming language (version 3.3 or newer). For pip\footnote{The Python package manager pip is handy when it comes to installing Python packages.} to be installed as part of Python, Python needs to be built against the Open\textsc{ssl} and zlib libraries. The Python package Blessings depends on the \_curses module, which is installed as part of Python provided that Python is built against the ncurses library. The following\footnote{Some of these depend on other Python packages which are not listed. If installing via pip, these additional Python packages will automatically be installed.} Python site-packages are required:
    \begin{description}
    	\item[\textbf{Blessings}] This package depends on the \_curses Python module.
    	\item[\textbf{Cython}] Version 0.22.0 or newer, though not 0.23.
    	\item[\textbf{Cython\_\textsc{gsl}}] This needs to be linked to the \textsc{gsl} library.
    	\item[\textbf{\textsc{h}{\smaller\smaller 5}\textsc{p}y}] Version 2.4 or newer, linked to the \textsc{hdf}{\smaller\smaller 5} library and configured to be parallel.
    	\item[\textbf{Matplotlib}] This package depends on the FreeType library.
    	\item[\textbf{\textsc{mpi}{\smaller\smaller 4}\textsc{p}y}] Version 1.3.1 or newer, linked to the \textsc{mpi} library.
    	\item[\textbf{NumPy}]
    	\item[\textbf{Pexpect}]
    \end{description}
    \item[ImageMagick] This is only needed for producing renders when running with more than 1 \textsc{cpu} and can be omitted. ImageMagick needs to be built with png and zlib features, which require the pnglib and zlib libraries.
    \item[\textbf{\textsc{gadget}}] The GAlaxies with Dark matter and Gas intEracT $N$-body code (version 2.0.7). This is only needed for running some tests and can be omitted. \textsc{Gadget} is also dependent on \textsc{gsl}, \textsc{mpi} and \textsc{fftw}. Note that \textsc{gadget} is incompatible with \textsc{fftw} 3.x, so a separate \textsc{fftw} 2.x must also be installed.
\end{description}

\section{The \texttt{installer}}
The \textsc{co\textsl{n}cept} code and all of the dependencies listed in the previous section (including the dependencies of the dependencies \dots) can be installed and linked together automatically by running the \texttt{installer} Bash script. You can find the script in the GitHub repository\footnote{\url{https://raw.githubusercontent.com/jmd-dk/concept/master/installer}}, or you can simply run\footnote{The \texttt{bit.ly/concept\_nbody} part refers to the \textsc{url} \url{http://bit.ly/concept_nbody}, which redirects to the \texttt{installer} script on GitHub. If the short version does not work, use the full \textsc{url}.}\footnote{The square bracket indicates an optional argument.}
\begin{lstlisting}[language=bash, backgroundcolor=\color{yellow!20}, xleftmargin=\parindent, xrightmargin=\parindent, basicstyle=\small\ttfamily, columns=fullflexible, keepspaces=true]
bash <(wget -O- --no-ch bit.ly/concept_nbody) [install_dir]
\end{lstlisting}
to automatically fetch and execute the script. Here, \texttt{install\_dir} is the path to the directory in which \concept should be installed. If you leave out this path, you will be prompted for it. Naturally, it is also possible to save the \texttt{installer} to disk and invoke it like e.g.\
\begin{lstlisting}[language=bash, backgroundcolor=\color{yellow!20}, xleftmargin=\parindent, xrightmargin=\parindent, basicstyle=\small\ttfamily, columns=fullflexible, keepspaces=true]
./installer [install_dir]
\end{lstlisting}
The \texttt{installer} script will use \textsc{mpich} for the needed \textsc{mpi} implementation\footnote{Note on possible problems related to using \textsc{mpich} on clusters with several nodes: \textsc{MPICH} uses ssh to launch \textsc{mpi}-threads on remote nodes, which require that password-less login to the cluster nodes is enabled (this is not to be confused with password-less login to the cluster itself). Should you encounter problems when running multi-node \concept jobs, this may be the problem. In addition, newer version of ssh blocks password-less login if your home-directory has group write access.}. Besides the listed Python packages and their own Python package dependencies, the \texttt{installer} will install the pip and Yolk3k Python packages, which is used to install the primary packages.

The \texttt{installer} script will install \concept and the entire dependency stack within the chosen installation directory. Uninstalling everything therefore simply amounts to removing this directory.

On modern hardware, it takes the \texttt{installer} around \SIrange{1}{2}{hours} to complete the installation. This is mainly because the \texttt{installer} script insists on installing everything from source, rather than using pre-compiled binaries. One benefit from this choice is that the installed software becomes tailor-made for the hardware in question, optimizing the performance. The primary reason for this choice is however portability (e.g.\ the \texttt{installer} is able to install \concept and the dependency stack on both 32-bit and 64-bit architectures).

A significant fraction of the installation time is spent running the test suites that the many programs provide. To speed up the installation (and leave potential errors unnoticed!), you may skip these tests by invoke the \texttt{installer} with the \texttt{{-}{-}fast} option:
\begin{lstlisting}[language=bash, backgroundcolor=\color{yellow!20}, xleftmargin=\parindent, xrightmargin=\parindent, basicstyle=\small\ttfamily, columns=fullflexible, keepspaces=true]
# from local copy of installer
./installer --fast [install_dir]
\end{lstlisting}
\begin{lstlisting}[language=bash, backgroundcolor=\color{yellow!20}, xleftmargin=\parindent, xrightmargin=0.3\parindent, basicstyle=\small\ttfamily, columns=fullflexible, keepspaces=true]
# from the installer on GitHub
bash <(wget -O- --no-ch bit.ly/concept_nbody) --fast [install_dir]
\end{lstlisting}
To decrease the installation time further, the \texttt{{-}{-}fast} option also sets any idle time to zero, decreasing the chances for successful downloading for questionable connections.

\subsection{Using Pre-installed Libraries}
By default, any pre-installed versions of the above software will be ignored. Should you wish the \texttt{installer} to use such pre-installed components, you have to declare their directory paths through \texttt{\textit{name}\_dir} environment variables, where \texttt{\textit{name}} is the name\footnote{The valid \texttt{\textit{name}}s are exactly those of the dependencies listed in section \ref{sec:dependencies}, written in all lowercase. For the \textsc{fftw} 2.x library used by \textsc{gadget}, use \texttt{fftw\_for\_gadget\_dir}.} of the pre-installed program. E.g.\ to use a pre-installed \textsc{gsl} library, invoke the installer like so:
\begin{lstlisting}[language=bash, backgroundcolor=\color{yellow!20}, xleftmargin=\parindent, xrightmargin=\parindent, basicstyle=\small\ttfamily, columns=fullflexible, keepspaces=true, mathescape]
# from local copy of installer
gsl_dir=/path/to/gsl ./installer [$\cdots$]
\end{lstlisting}
\begin{lstlisting}[language=bash, backgroundcolor=\color{yellow!20}, xleftmargin=\parindent, xrightmargin=-1.3\parindent, basicstyle=\small\ttfamily, columns=fullflexible, keepspaces=true, mathescape]
# from the installer on GitHub
gsl_dir=/path/to/gsl bash <(wget -O- --no-ch bit.ly/concept_nbody) [$\cdots$]
\end{lstlisting}
Here and in what follows, \verb|[|$\cdots$\verb|]| stands for a space-seperated list of optional arguments (e.g.\ \texttt{install\_dir} and \texttt{{-}{-}fast}). Multiple program paths are specified as a space-separated list.

Should you choose to use a pre-installed Python distribution, it needs to satisfy the following requirements:
\begin{itemize}
\item The pip Python package needs to be installed if any of the required Python packages are not installed.
\item The \_curses Python module needs to be installed.
\end{itemize}
Pre-installed Python packages are automatically detected when the \texttt{python\_dir} environment variable is set. That is, you do not have to set e.g.\ \texttt{blessings\_dir}.

\subsection{Requesting Specific Versions}
Should you wish to install a specific version of some dependency, simply set the environment variable \texttt{\textit{name}\_version} when invoking the installer. This also works for Python packages. E.g.\ to install version 1.15 of \textsc{gsl}:
\begin{lstlisting}[language=bash, backgroundcolor=\color{yellow!20}, xleftmargin=\parindent, xrightmargin=\parindent, basicstyle=\small\ttfamily, columns=fullflexible, keepspaces=true, mathescape]
# from local copy of installer
gsl_version=1.15 ./installer [$\cdots$]
\end{lstlisting}
\begin{lstlisting}[language=bash, backgroundcolor=\color{yellow!20}, xleftmargin=\parindent, xrightmargin=-0.1\parindent, basicstyle=\small\ttfamily, columns=fullflexible, keepspaces=true, mathescape]
# from the installer on GitHub
gsl_version=1.15 bash <(wget -O- --no-ch bit.ly/concept_nbody) [$\cdots$]
\end{lstlisting}
Multiple program versions are specified as a space-separated list.

\section{Manual Installation}
To do a manual installation of any of the dependency programs, we refer to their individual online documentation.

To install just the \concept code without any of the dependency programs, simply download the \concept code from its GitHub repository\footnote{\url{https://github.com/jmd-dk/concept/archive/master.tar.gz}} and extract it into some directory. To make \concept aware of the installed dependency programs, simply edit the \texttt{.paths} file, which you will find in the topmost directory. The \texttt{.paths} file contains the \emph{absolute} paths to every\footnote{With the exception of system dependencies (e.g.\ \textsc{awk}) which are expected to be on the \texttt{PATH}.} file and directory which \concept needs to know about. No path information is ever drawn from anywhere else. The description of each variable within the \texttt{.paths} file itself should constitute enough documentation of which variable should contain which path.

All the paths specified in the \texttt{.paths} file are read in at runtime by \concept. Should you ever wish to make \concept aware of additional paths, simply add them to the \texttt{.paths} file.

\chapter{Running the Code}
In this chapter the primary functionalities and invocation methods of \concept are described. In addition, \concept implements secondary functionalities --- called \emph{utilities} --- which are not used to run simulations directly, but serve as convenient tools within the program environment. For these secondary uses of \concept, see chapter \ref{ch:utilities}.

\begin{figure}[tb]
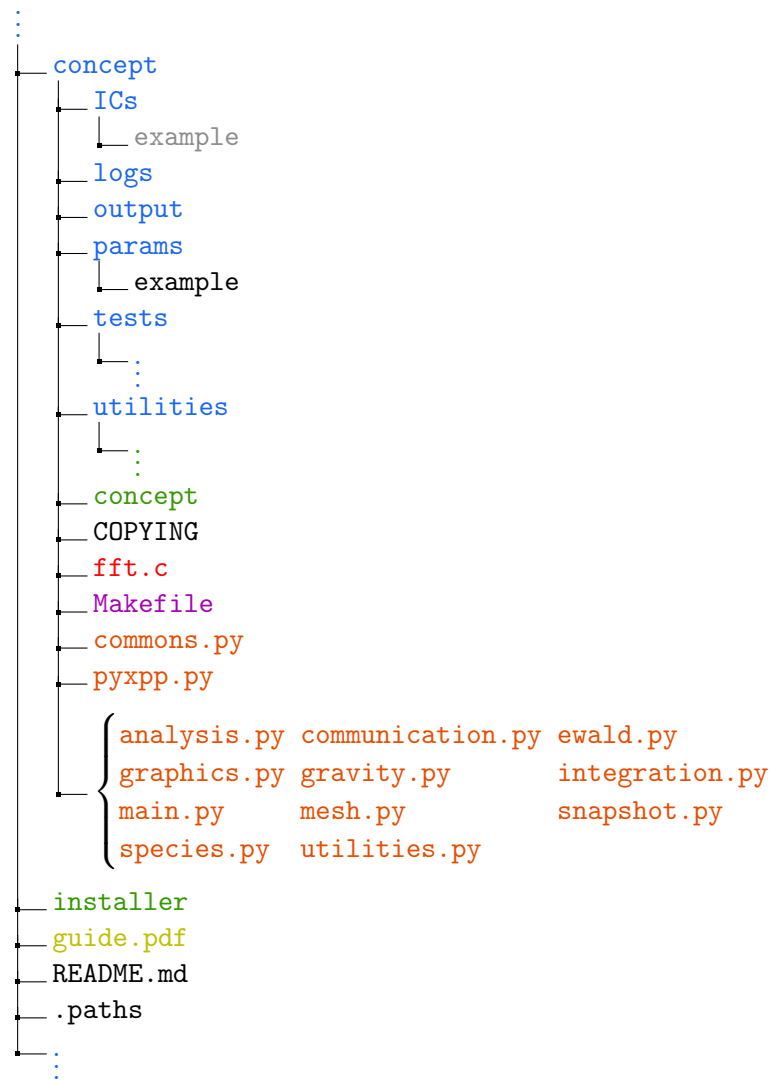

\center
\begin{minipage}{0.8\linewidth}
\dirtree{%
.1 \hspace{3pt}\smash{\filedir{$\vdots$}}.
.2 \filedir{concept}.
.3 \filedir{ICs}.
.4 \filehdf{example}.
.3 \filedir{logs}.
.3 \filedir{output}.
.3 \filedir{params}.
.4 \filetext{example}.
.3 \filedir{tests}.
.4 \raisebox{-0.84em}{\filedir{$\vdots$}}.
.3 \filedir{utilities}.
.4 \raisebox{-0.84em}{\filebash{$\vdots$}}.
.3 \filebash{concept}.
.3 \filetext{COPYING}.
.3 \filec{fft.c}.
.3 \filemakefile{Makefile}.
.3 \filepy{commons.py}.
.3 \filepy{pyxpp.py}
\vspace{27pt}.
.3 \partsmash[t]{0.34}{$\begin{cases}
\text{\filepy{analysis.py}{ }\filepy{communication.py}{ }\filepy{ewald.py}}\\[-2pt]
\text{\filepy{graphics.py}{ }\filepy{gravity.py}{ }{ }{ }{ }{ }{ }{ }\filepy{integration.py}}\\[-2pt]
\text{\filepy{main.py}{ }{ }{ }{ }{ }\filepy{mesh.py}{ }{ }{ }{ }{ }{ }{ }{ }{ }{ }\filepy{snapshot.py}}\\[-2pt]
\text{\filepy{species.py}{ }{ }\filepy{utilities.py}}
\end{cases}$}
\vspace{-10pt}.
.2 \filebash{installer}.
.2 \filepdf{guide.pdf}.
.2 \filetext{README.md}.
.2 \filetext{.paths}.
.2 \raisebox{-0.84em}{\filedir{$\vdots$}}.
}
\end{minipage}
\caption{Directory tree for \concept. As in figure \ref{fig:basicdirtree}, the colors indicate directories (blue), Bash scripts (green), text files (black) and \textsc{hdf}{\smaller 5} files (grey). In addition we now have C source code files (red), makefiles (purple), \textsc{pdf} files (yellow) and Python files (orange). Two of the Python files are listed separately, while the others are grouped together. The files in this group are referred to as Python \emph{modules}.}
\label{fig:dirtree}
\end{figure}

A directory tree of \concept is shown in figure \ref{fig:dirtree}. Disregarding IC and parameter files, the \texttt{concept} script is the only file that the user should ever have to interact with directly in order to run the code. In the simplest case (see the next section for the differences between local and remote execution), executing the \texttt{concept} script will build and run the code.

As compilation is performed automatically when the code is executed, before describing how to run the code in detail, a brief description of this build process is in order. When the \texttt{concept} script is executed, it invokes the \texttt{Makefile} which builds the program. The compiled program consists of a set of shared object (\texttt{.so}) files; precisely one for each of the modules (the Python \texttt{.py} files grouped together in figure \ref{fig:dirtree}). To run the program, the \texttt{main} module, \texttt{main.so}, is invoked via the Python interpreter, which itself is started by \textsc{mpi}. Running in so-called \quote{pure Python mode}, meaning without first having compiled the code, amounts to skipping the entire build process but otherwise invoking the \texttt{main} module (now \texttt{main.py}) in a similar fashion. For the exact commands needed to run \concept manually, see section \ref{manualbuildrun}.

Each \concept run will be assigned a unique ID. The combined output to \texttt{stdout} and \texttt{stderr} will always be logged to the file \texttt{logs/\textit{ID}}, where \texttt{\textit{ID}} is the ID of the current run in question. The output from \texttt{stderr} alone will similarly be saved to \texttt{logs/\textit{ID}\_err}.

\section{Local vs.\ Remote Execution}
Running the \texttt{concept} script on a local machine will build (if necessary) and run the code. When logged on to a remote host though, the default behavior is to build the program and do nothing more. This behavior is chosen in order to bring the user in control of how the program should be executed. Often, computer clusters utilizes a job scheduling mechanism, such as \textsc{pbs} (Portable Batch System). The \concept script \emph{is} able to autogenerate and submit \textsc{pbs} job scripts\footnote{Even though several versions of \textsc{pbs} exist (such us Open\textsc{pbs}, \textsc{torque} and \textsc{pbs} Professional), each with slightly different functionality, the core components used by \texttt{concept} are the same in all of them.}. If this feature is desired, you need to edit the first couple of lines in the \texttt{concept} script as follows:
\begin{itemize}
	\item[]{Populate the \texttt{queues} and \texttt{ppns} variables with queue names and corresponding numbers of \textsc{cpu}s per node. As an example, imagine that we wish \concept to be aware of the queues {\lq}q8{\rq} and {\lq}q4{\rq}, which have 8 and 4 \textsc{cpu}s per node, respectively. In that case, we would set the variables as\\
	\texttt{queues=(q8 q4)}\\
	\texttt{ppns=(8 4)}}
\end{itemize}
Now when executing the \texttt{concept} script on the remote host, a queue from the \texttt{queues} variable will be chosen for the job, based on its number of \textsc{cpu}s per node. To ensure maximum exploitation of the available resources, a queue will only be deemed fit for use if the number of requested processes is divisible by the number of \textsc{cpu}s per node in the queue (so that the job utilizes every \textsc{cpu} on the nodes it occupy). If no queue satisfies this requirement, the job will not be submitted. If multiple queues satisfy the requirement, the one listed first in the \texttt{queues} variable will be used. A job script will then be generated, saved as \texttt{concept/jobscript} and submitted.

Should you wish to run \concept directly, without the use of \textsc{pbs}, even though you are not working on a local machine, invoke \texttt{concept} with the \texttt{{-}{-}local} option:
\begin{lstlisting}[language=bash, backgroundcolor=\color{yellow!20}, xleftmargin=\parindent, xrightmargin=\parindent, basicstyle=\small\ttfamily, columns=fullflexible, keepspaces=true, mathescape]
./concept --local
\end{lstlisting}
The next section lists with the full set of options available when executing \texttt{concept}.

\section{The \texttt{concept} Command-Line Interface}\label{sec:commandlineinterface}
As already demonstrated, several optional options may be given when invoking the \texttt{concept} script. As is customary for \textsc{unix} shell scripts, a list of available options together with brief descriptions will be shown when invoking the \texttt{concept} script with the \texttt{-h} or \texttt{{-}{-}help} option. Here follows the same list but with much extended descriptions. Remember that supplied paths may be either absolute or relative to the \texttt{concept} directory.
\begin{description}[font=\normalfont\ttfamily]
    \item[-h, {-}{-}help] \hfill \\
    Show the help message and exit.
    \item[-m MAIN, {-}{-}main MAIN] \hfill \\
    Sets the main entry point of the code to the path \texttt{MAIN}. The default value is \texttt{concept/main.py}. This option is used internally by \concept when running tests, but does not currently have other use cases.
    \item[-n NPROCS, {-}{-}nprocs NPROCS] \hfill \\
	Sets the number of \textsc{mpi} processes to \texttt{NPROCS}. The default value is 1. Note that this can take on values larger than the number of available \textsc{cpu}s, at the cost of performance.
    \item[-p PARAMS, {-}{-}params PARAMS] \hfill \\
    Sets the parameter file to be used to \texttt{PARAMS}. If not set, default parameters as specified in \texttt{commons.py} are used.
    \item[-q QUEUE, {-}{-}queue QUEUE] \hfill \\
	Sets the queue to be used when submitting as a remote \textsc{pbs} job to \texttt{QUEUE}. The specified queue must be predefined in the \texttt{queues} and \texttt{ppns} variables in the top of the \texttt{concept} script. If this option is not set, the first queue in the \texttt{queues} variable which have a matching number of \textsc{cpu}s per node will be chosen.
    \item[-t TEST, {-}{-}test TEST] \hfill \\
    Instead of running a simulation, run the test \texttt{TEST}. The possible values for \texttt{TEST} are the names of the subdirectories within the \texttt{concept/tests} directory, possibly with the relative or absolute path included. In addition it is possible to run all tests by setting \texttt{TEST} to \texttt{all}. See chapter \ref{ch:tests} for details.
    \item[-u UTIL {[}$\cdots${]}, {-}{-}util UTIL {[}$\cdots${]}] \hfill \\
    Instead of running a simulation, run the utility \texttt{UTIL}. The possible values for \texttt{UTIL} are the names of the subdirectories within the \texttt{concept/utilities} directory, possibly with the relative or absolute path included. Most utilities require an additional argument to be passed. See chapter \ref{ch:utilities} for details.
	\item[-w WALLTIME, {-}{-}walltime WALLTIME] \hfill \\
	Sets the \textsc{pbs} walltime to \texttt{WALLTIME}, in whole hours. The default is 72 hours.
    \item[{-}{-}local] \hfill  \\
    Build and run the code as if working on a local machine, regardless of whether this is actually the case. That is, even if working remotely, the code will be executed directly, rather than submitted as a \textsc{pbs} job.
    \item[{-}{-}pure-python] \hfill \\
    Skips the build process. The code will then resort to use the raw Python files directly. This is generally not preferred due to the severe performance penalty this entails, but it can be useful for debugging and running the code with a main entry point not suited for cythonization. If the code is already fully built, the shared object (\texttt{.so}) files will be renamed to \texttt{.so\_}, in order for Python not to pick up these compiled modules. The next time \texttt{concept} is invoked without the \texttt{{-}{-}pure-python} option, the compiled module files will be renamed back to \texttt{.so}.
\end{description}

\section{Manual Build and Run}\label{manualbuildrun}
Should you ever need to build and run \concept manually, i.e.\ without invoking the \texttt{concept} script, this section lists the needed commands. When running the code manually, logging is your responsibility.

To build the code, simply type
\begin{lstlisting}[language=bash, backgroundcolor=\color{yellow!20}, xleftmargin=\parindent, xrightmargin=\parindent, basicstyle=\small\ttfamily, columns=fullflexible, keepspaces=true, mathescape]
make
\end{lstlisting}
within the \texttt{concept} directory. When the code is fully built, you can execute it as

\begin{lstlisting}[language=bash, backgroundcolor=\color{yellow!20}, xleftmargin=\parindent, xrightmargin=\parindent, basicstyle=\small\ttfamily, columns=fullflexible, keepspaces=true, mathescape]
mpiexec -n NPROCS python -B -m main.so "params='PARAMS'"
\end{lstlisting}
where \texttt{mpiexec} and \texttt{python} are, respectively, the \textsc{mpi} executable and the Python interpreter, the paths of which should be written in the \texttt{.paths} file. As in section \ref{sec:commandlineinterface}, \texttt{NPROCS} and \texttt{PARAMS} are the number of processes and the path to the parameter file, respectively.

Should you wish to run in pure Python mode, you first have to remove the shared object files. To remove all files resulting from compilation, type
\begin{lstlisting}[language=bash, backgroundcolor=\color{yellow!20}, xleftmargin=\parindent, xrightmargin=\parindent, basicstyle=\small\ttfamily, columns=fullflexible, keepspaces=true, mathescape]
make clean
\end{lstlisting}
Now to run \concept in pure Python mode, execute is as
\begin{lstlisting}[language=bash, backgroundcolor=\color{yellow!20}, xleftmargin=\parindent, xrightmargin=\parindent, basicstyle=\small\ttfamily, columns=fullflexible, keepspaces=true, mathescape]
mpiexec -n NPROCS python -B main.py "params='PARAMS'"
\end{lstlisting}

\section{Cleaning}
Sometimes it is useful to be able to quickly remove a particular set of files. Therefore the \texttt{Makefile} defines the following clean targets:
\begin{description}
	\item[clean] \hfill \\
	Removes all files created by the build process.
	\item[clean\_auxiliary] \hfill \\
	Removes files generated by running \concept itself, like the \texttt{jobscript} and \textsc{fftw} wisdom.
	\item[clean\_logs] \hfill \\
	Removes all files in the \texttt{logs} directory.
	\item[clean\_output] \hfill \\
	Removes all files and directories in the \texttt{output} directory.
	\item[clean\_tests] \hfill \\
	Removes all files in the subdirectories of the \texttt{test} directory, which were created by running tests.
	\item[clean\_utilities] \hfill \\
	Remove all files generated as a bi-product of using utilities.
	\item[distclean] \hfill \\
	Calls all of the targets above. If you yourself have not created any additional files, the \concept environment is now in a distribution ready state.
\end{description}

\chapter{Parameters}\label{ch:params}
The parameters of a given run must be supplied in a parameter file, which is just a text file containing Python assignments. An example of such a parameter file is \texttt{concept/params/example}. In it, a short description of each parameter is written next to it. A complete description of each parameter is given below.

When assigning values to a parameter with some physical unit, the user is strongly urged to write this unit explicitly in the value. As an example consider the specification of \texttt{boxsize}, the linear size of the simulation box:
\begin{lstlisting}[language=bash, xleftmargin=\parindent, xrightmargin=\parindent, basicstyle=\small\ttfamily, columns=fullflexible, columns=fullflexible, keepspaces=true, mathescape]
boxsize = 100*Mpc  # Great!
boxsize = 100      # Will probably work, but bad!
\end{lstlisting}
Here, the first definition is much preferable to the latter, even though \concept by default uses \si{Mpc} as the base unit for lengths, and so numerically the definitions are equivalent. For one, having implicit units floating around is not good practice. More importantly, it will lead to trouble if someone were to redefine the internal unit system used by \concept, which can easily be done by changing a few lines in the \texttt{Units} class in \texttt{commons.py}. Should you wish to extend the set of units known by \concept, this is also the place to go.

In addition to physical units, any valid Python construct are allowed in parameter definitions. All normal mathematical as well as NumPy functions and constants are also available. As always, paths may be absolute or relative to the \texttt{concept} directory. In addition, paths from the \texttt{.paths} file are accessible as e.g.\ \texttt{paths['concept\_dir']} for the \texttt{concept} directory. The directory of the parameter file itself is accessible as \texttt{paths['params\_dir']}.

Here follows a full list of parameters, grouped into five sections. 
\begin{description}
    \item[Input/output] \hfill \vspace{-0.4em}
    \begin{description}[font=\normalfont\ttfamily]
        \item[IC\_file] \hfill \\
        Path to the snapshot file containing the initial conditions for the simulation. The snapshot may be in any format known to \concept. The code will automatically figure out the snapshot format at load time.
        \item[snapshot\_type] \hfill \\
        The format of the output snapshots. The implemented snapshot formats are described in detail in chapter \ref{ch:snapshots}. At the time of writing, \concept implements two snapshot formats, called \texttt{'standard'} and \texttt{'GADGET 2'}, referring to the code's own \textsc{hdf}{\smaller\smaller 5} snapshot type and the secondary snapshot type defined by \textsc{gadget-}{\smaller\smaller 2}, respectively.
        \item[output\_dirs] \hfill \\
        This is a dictionary with keys (left of colon) corresponding to output types and values (right of colon), giving the paths to the respective output directories. That is, its general form is like the following:
\begin{lstlisting}[language=bash, xleftmargin=\parindent, xrightmargin=\parindent, basicstyle=\small\ttfamily, columns=fullflexible, columns=fullflexible, keepspaces=true, mathescape]
output_dirs = {
    'snapshot' : '/path/to/snapshots',
    'powerspec': '/path/to/powerspecs',
    'render'   : '/path/to/renders',
               }
\end{lstlisting}
        Output directories of snapshots, power spectra and renders are thus defined individually. If a specified directory does not exist, it will be created.                
        \item[output\_bases] \hfill \\
        This is a dictionary of the same form as \texttt{output\_dirs}. Its values refer to the basenames (prefixes) of the output files.
         \item[output\_times] \hfill \\
         This is a dictionary of the same form as \texttt{output\_dirs}, though with the additional \texttt{'terminal render'} key. The values correspond to those values of the scale factor for which an output should be produced. A value can be a single number or any Python sequence of numbers. Leave a value empty (set it to e.g.\ \texttt{''}) or remote its key completely in order not to produce output of this kind.
    \end{description}
    \item[Numerical parameters] \hfill \vspace{-0.4em}
    \begin{description}[font=\normalfont\ttfamily]
        \item[boxsize] \hfill \\
        The linear, comoving size of the cubic simulation box. The simulated, comoving volume is then $\texttt{boxsize}^3$.
        \item[ewald\_gridsize] \hfill \\
        Linear size of the grid of Ewald corrections. The total number of tabulated Ewald corrections is then $\texttt{ewald\_gridsize}^3$.
        \item[PM\_gridsize] \hfill \\
        Linear size of the grid used by the PM method as well as the power spectrum computation. The simulation volume will then be discretized into $\texttt{PM\_gridsize}^3$ grid points.        
        \item[PM\_scale] \hfill \\       
        The scale of the gravitational force splitting into a long-range component and a short-range component, used in the $\textsc{p}^{3}\textsc{m}$ method. The scale should be given in units of the linear size of a PM cell. A value slightly above 1 should be optimal.
        \item[P3M\_cutoff] \hfill \\  
        Distance beyond which short-range forces are ignored in the $\text{P}^{3}\text{M}$ method. This distance should be given in units of \texttt{P3M\_scale}. A value of at least 4.8 guarantees that the force components left unaccounted for constitutes less than \SI{1}{\percent} of the total force.
        \item[softeningfactors] \hfill \\
        Mapping from particle species to values of their respective gravitational softening. The values given correspond to the radius of a softening Plummer sphere, in units of the mean inter-particle distance $\texttt{boxsize}/N^{\nicefrac{1}{3}}$. For dark matter, a value of a couple of percent is optimal.
        \item[$\mathrm{\Delta}$t\_factor] \hfill \\
        The global time step size, given in units of the instantaneous age of the universe.
        \item[R\_tophat] \hfill \\
        Radius of the top-hat used to compute $\sigma$, the rms density variation, from the power spectrum.
    \end{description}
    \item[Cosmological parameters] \hfill \vspace{-0.4em}
    \begin{description}[font=\normalfont\ttfamily]
        \item[H0] \hfill \\
        The Hubble parameter at the present time ($a = 1$). If this does not match the corresponding value in the \texttt{IC\_file}, a warning will be thrown.
        \item[$\mathrm{\Omega}$m] \hfill \\
        The matter density parameter at the present time ($a = 1$). If this does not match the corresponding value in the \texttt{IC\_file}, a warning will be thrown.
        \item[$\mathrm{\Omega\Lambda}$] \hfill \\
        The dark energy density parameter at the present time ($a = 1$). If this does not match the corresponding value in the \texttt{IC\_file}, a warning will be thrown.
        \item[a\_begin] \hfill \\
        The value of the scale factor at the beginning of the simulation. If this does not match the value of the scale factor in the \texttt{IC\_file}, a warning will be thrown.
    \end{description}
    \item[Graphics] \hfill \vspace{-0.4em}
    \begin{description}[font=\normalfont\ttfamily]
        \item[powerspec\_plot] \hfill \\
        Boolean value determining whether or not when outputting a power spectrum (a text file of data), a plot of the data should also be made.
    \item[color] \hfill \\
    The color of the particles in the renders. This can be a number in the range 0--1 (grayscale), any sequence of 3 numbers in the range 0--1 (\textsc{rgb} values) or a string containing any of the color names defined by Matplotlib. See \url{http://matplotlib.org/mpl_examples/color/named_colors.hires.png} for the entire repertoire.
    \item[bgcolor] \hfill \\
    The background color on the renders. The format is similar to that of \texttt{color}.
    \item[resolution] \hfill \\
    The height and width of the saved renders, in pixels. Note that the images will always be square.
        \item[liverender] \hfill \\
        Path to the file where the latest frame should be saved, resulting in a \quote{live} visualization of the simulation. Set an empty path to disable this feature.
    \item[remote\_liverender] \hfill \\
    If live render is used, this parameter allows for uploading of the live render to a remote host. Secure copy (scp) is used. The parameter takes the usual scp form; \texttt{'user@host:/path/to/liveframe'}. When using this parameter, \concept will prompt you once for your password to the host. Use an empty string to disable this feature.
    \item[terminal\_colormap] \hfill \\
    Sets the colormap used in terminal renders. This can be any colormap defined by Matplotlib. See \url{http://matplotlib.org/examples/color/colormaps_reference.html} for a complete list.
    \item[terminal\_resolution] \hfill \\
    The width of the terminal renders, in characters. The height is chosen to be half as big as the width, resulting in roughly rectangular (pixel-wise) terminal renders on modern terminals where each character is about twice as high as they are wide.
    \end{description}
    \item[Simulation options] \hfill \vspace{-0.4em}
    \begin{description}[font=\normalfont\ttfamily]
        \item[kick\_algorithms] \hfill \\
        Mapping from particle species to the algorithm which should be used when kicking particles of that species. Implemented kick algorithms are \texttt{'PP'}, \texttt{'PM'} and \texttt{'P3M'}.
        \item[use\_Ewald] \hfill \\
        This should normally be set to \texttt{True}, in which case periodic corrections to the gravitational force will be computed via Ewald summation, when using the PP method. Otherwise, no Ewald corrections will be computed.
        \item[fftw\_rigor] \hfill \\
        The \quote{rigor level} used when acquiring \textsc{fftw} wisdom. It can be any of \texttt{'estimate'}, \texttt{'measure'}, \texttt{'patient'} or \texttt{'exhaustive'}. For the official \textsc{fftw} documentation about each rigor level, see \url{http://www.fftw.org/doc/Planner-Flags.html}. If \textsc{fftw} wisdom of the right kind but for a higher level rigor than requested exists, this higher level rigor is used instead.
    \end{description}
\end{description}
Each parameter has an associated default value, defined in the \texttt{commons.py} file. If your run does not make use of every parameter, it may be desirable to remove those parameter definitions from the parameter file, leaving a more comprehensible file, only stating the definitions you actually care about. An example would be to remove the entire graphics category, if your run does not output any renders or terminal renders. It is however not advised to use these default values in situations where your run depend upon which value is given to some parameter.

\chapter{Snapshot Formats}\label{ch:snapshots}
At the time of writing, two snapshot formats are supported by \concept. One of them is simply referred to as the \quote{standard} \concept snapshot format, while the other is the \textsc{gadget-}{\smaller\smaller 2} format of the second kind\footnote{For details of the \textsc{gadget-}{\smaller\smaller 2} format of the second kind, see \url{http://wwwmpa.mpa-garching.mpg.de/gadget/users-guide.pdf} chapter 6.}. When using a \textsc{gadget-}{\smaller\smaller 2} snapshot with \concept, certain restrictions apply:
\begin{itemize}
	\item In the jargon of \textsc{gadget-}{\smaller\smaller 2}, dark matter particles are referred to as \quote{halo} particles or particles of type 1. As dark matter is the only implemented species in \concept, for a \textsc{gadget-}{\smaller\smaller 2} snapshot to be usable by \concept, it must consist exclusively of particles of type 1.
	\item \textsc{Gadget-}{\smaller\smaller 2} supports distributing a single snapshot across multiple files. This is not implemented in \concept, and so for a \textsc{gadget-}{\smaller\smaller 2} snapshot to be readable by \concept, the snapshot must be completely contained within a single file.
\end{itemize}
The reason for making \concept at least somewhat compatible with foreign snapshot types is to connect it with existing software, where the \textsc{gadget-}{\smaller\smaller 2} snapshot format are somewhat standard. The standard \concept snapshot format is however preferable, as it uses the well-established \textsc{hdf}{\smaller\smaller 5} format\footnote{As \textsc{gadget-}{\smaller\smaller 2} \emph{is} able to read and write snapshots in the \textsc{hdf}{\smaller\smaller 5} format, this is not a critique of \textsc{gadget-}{\smaller\smaller 2} itself.}.

\section{The Standard \concept Snapshot Format}\label{sec:standardsnapshot}
\begin{figure}[tb]
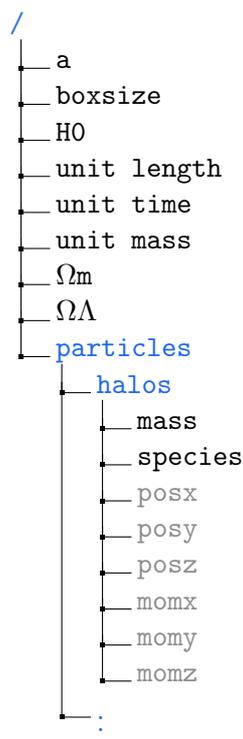

\center
\begin{minipage}{0.3\linewidth}
\dirtree{%
.1 \filedir{/}.
.2 a.
.2 boxsize.
.2 H0.
.2 unit length.
.2 unit time.
.2 unit mass.
.2 $\mathrm{\Omega}$m.
.2 $\mathrm{\Omega\Lambda}$.
.2 \filedir{particles}.
.3 \filedir{halos}.
.4 mass.
.4 species.
.4 \filehdf{posx}.
.4 \filehdf{posy}.
.4 \filehdf{posz}.
.4 \filehdf{momx}.
.4 \filehdf{momy}.
.4 \filehdf{momz}.
.3 \raisebox{-0.84em}{\filedir{$\vdots$}}.
}
\end{minipage}
\caption{Layout of the standard \textsc{hdf}{\smaller 5} snapshot format used by \concept. The colors indicate groups (blue), attributes (black) and datasets (gray).}
\label{fig:standardsnapshotlayout}
\end{figure}

\Concept's own snapshot format, the \quote{standard} format, is really just an \textsc{hdf}{\smaller\smaller 5} file with some particular layout. This layout is shown in figure \ref{fig:standardsnapshotlayout}. The root group contains all of the cosmological parameters also present in parameter files\footnote{Here, the scale factor is simply called \texttt{a} rather than \texttt{a\_begin}, as the snapshot is not necessarily intended as an IC file.}. This allows \concept to do sanity checks when reading in an IC file.

Beside attributes containing information about the cosmology, the root group also contains the three \texttt{unit}s attributes. Their values are strings telling which physical unit is used as the base unit, for length, time and mass. When a standard snapshot is read in by \concept, all physical quantities gets multiplied by the appropriate unit. With the unit information contained in the snapshot, the internal base unit of the current run may differ from that which produced the snapshot. This makes the snapshots more general and self-documenting.

Even with only a single implemented particle species, one may wish to use several groups of particles, which differ in e.g.\ mass. At the time of writing, the \concept code is not capable of reading in snapshots containing different particles, but the structure of the snapshots allows it. Each group of particles gets a distinctive label called its \emph{type}. In figure \ref{fig:standardsnapshotlayout}, the only group within the root group is \texttt{particles}, which in turn stores all particle types as subgroups. In figure \ref{fig:standardsnapshotlayout}, \texttt{halo} is shown as an example of a particle type. This group contains a \texttt{mass} attribute, giving the mass of each particle of the halo type in units of \texttt{unit mass}, together with a \texttt{species} attribute, which currently must equal \texttt{'dark matter'}. The particle positions and momenta are stored as 6 datasets; \texttt{posx}, \texttt{posy}, \texttt{posz} for the position components and \texttt{momx}, \texttt{momy}, \texttt{momz} for the momenta components. Note that since these datasets know their own size, the number of particles is not saved as an attribute. The positions are given in units of \texttt{units length}, while the momenta are given in units of $(\texttt{unit length})\times(\texttt{unit mass})/(\texttt{unit time})$. Naturally, these are the \emph{comoving} positions and momenta.

\chapter{Utilities}\label{ch:utilities}
\begin{figure}[tb]
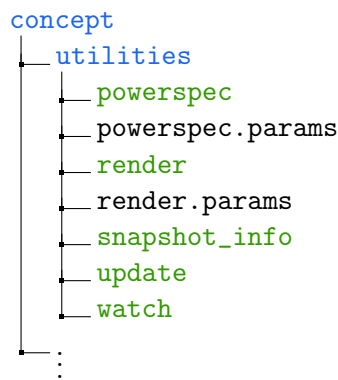

\center
\begin{minipage}{0.3\linewidth}
\dirtree{%
.1 \filedir{concept}.
.2 \filedir{utilities}.
.3 \filebash{powerspec}.
.3 \filetext{powerspec.params}.
.3 \filebash{render}.
.3 \filetext{render.params}.
.3 \filebash{snapshot\_info}.
.3 \filebash{update}.
.3 \filebash{watch}.
.2 \raisebox{-0.84em}{\filetext{$\vdots$}}.
}
\end{minipage}
\caption{Directory tree focusing on the \texttt{utilities} directory. The color codes are identical to those of figure \ref{fig:dirtree}.}
\label{fig:utildirtree}
\end{figure}

The content of \texttt{concept/utilities} is missing from figure \ref{fig:dirtree}. This is instead shown in figure \ref{fig:utildirtree}. Each Bash script is its own utility, capable of performing a small task. These should be called via the \texttt{concept} script as described in section \ref{sec:commandlineinterface}, i.e.\ via the \texttt{-u} option. Here follows a description of each utility.

\begin{description}[font=\normalfont\ttfamily]
    \item[powerspec] \hfill \\
    This utility uses \concept to compute power spectra of particles in snapshots. Invoke it as
\begin{lstlisting}[language=bash, backgroundcolor=\color{yellow!20}, xleftmargin=\parindent, xrightmargin=\parindent, basicstyle=\small\ttfamily, columns=fullflexible, keepspaces=true, mathescape]
./concept -u powerspec PATH
\end{lstlisting}
    where \texttt{PATH} is a path to either a snapshot or a directory containing (maybe among other files) at least one snapshot. All the usual options of the \texttt{concept} script still apply. If no parameter file is specified, \texttt{powerspec.params} is used.
    \item[render] \hfill \\
    This utility uses \concept to produce renders of particles in snapshots. Invoke it as
\begin{lstlisting}[language=bash, backgroundcolor=\color{yellow!20}, xleftmargin=\parindent, xrightmargin=\parindent, basicstyle=\small\ttfamily, columns=fullflexible, keepspaces=true, mathescape]
./concept -u render PATH
\end{lstlisting}
    where \texttt{PATH} is a path to either a snapshot or a directory containing (maybe among other files) at least one snapshot. All the usual options of the \texttt{concept} script still apply. If no parameter file is specified, \texttt{render.params} is used.
    \item[snapshot\_info] \hfill \\
    This utility will print out the information contained within snapshots. Invoke it as
\begin{lstlisting}[language=bash, backgroundcolor=\color{yellow!20}, xleftmargin=\parindent, xrightmargin=\parindent, basicstyle=\small\ttfamily, columns=fullflexible, keepspaces=true, mathescape]
./concept -u snapshot_info PATH
\end{lstlisting}
    where \texttt{PATH} is a path to either a snapshot or a directory containing (maybe among other files) at least one snapshot. This utility will always be run locally in pure Python mode, using a single \textsc{cpu}.	For standard snapshots, it will simply print out the values of the attributes described in section \ref{sec:standardsnapshot}, along with the number of particles and their type. The same information is stored in \textsc{gadget-}{\smaller\smaller 2} snapshots, but in very different units. When using the \texttt{snapshot\_info} utility on a \textsc{gadget-}{\smaller\smaller 2} snapshot, its attribute values will be converted to the \quote{standard} form before being printed out. Comparing a standard snapshot to a \textsc{gadget-}{\smaller\smaller 2} is thus made easy, as all the unit conversions are being taken care of. A \textsc{gadget-}{\smaller\smaller 2} snapshot also contains a lot of other, \textsc{gadget-}{\smaller\smaller 2} specific parameters. These will also get printed out, in their raw form.
    \item[update] \hfill \\
    This utility will update all of the \concept source files to their latest versions, including the \texttt{update} script itself. Invoke it either as
\begin{lstlisting}[language=bash, backgroundcolor=\color{yellow!20}, xleftmargin=\parindent, xrightmargin=\parindent, basicstyle=\small\ttfamily, columns=fullflexible, keepspaces=true, mathescape]
./concept -u update
\end{lstlisting}
	or directly as
\begin{lstlisting}[language=bash, backgroundcolor=\color{yellow!20}, xleftmargin=\parindent, xrightmargin=\parindent, basicstyle=\small\ttfamily, columns=fullflexible, keepspaces=true, mathescape]
utilities/update
\end{lstlisting}
Needless to say, this utility will always be run locally using a single \textsc{cpu}. Before actually doing the update, a backup of the original \concept code will be placed in \texttt{../concept\_backup}.
    \item[watch] \hfill \\
    This script will continuously print out the output of a \concept run, after it has been submitted as a \textsc{pbs} job. It does this by reading the produced logfile. If you have multiple jobs submitted, the one submitted most recently will be chosen. Alternatively, the \textsc{pbs} job ID can be passed as an argument when calling the script. That is, call this script either as 
\begin{lstlisting}[language=bash, backgroundcolor=\color{yellow!20}, xleftmargin=\parindent, xrightmargin=\parindent, basicstyle=\small\ttfamily, columns=fullflexible, keepspaces=true, mathescape]
./concept -u watch [ID]
\end{lstlisting}
	or directly as
\begin{lstlisting}[language=bash, backgroundcolor=\color{yellow!20}, xleftmargin=\parindent, xrightmargin=\parindent, basicstyle=\small\ttfamily, columns=fullflexible, keepspaces=true, mathescape]
utilities/watch [ID]
\end{lstlisting}
Needless to say, this utility will always be run locally using a single \textsc{cpu}. After \concept auto-submits a \textsc{pbs} job, this script is automatically called.
\end{description}

\chapter{Tests}\label{ch:tests}
Code validation is achieved through unit tests. Each subdirectory within the \texttt{tests} constitutes such a test, the name of which is equal to the directory name. To run e.g.\ the \texttt{basic} test, invoke \texttt{concept} like this:
\begin{lstlisting}[language=bash, backgroundcolor=\color{yellow!20}, xleftmargin=\parindent, xrightmargin=\parindent, basicstyle=\small\ttfamily, columns=fullflexible, keepspaces=true, mathescape]
./concept -t basic
\end{lstlisting}

Each test is controlled by a Bash script named \texttt{run\_test}, which (with the help of other files in the same directory) runs the \textsc{co\textsl{n}cept} code with a certain set of parameters, analyses the output and report any problems. Often, suitable initial conditions are generated for the test, which are then evolved by both \textsc{co\textsl{n}cept} and \textsc{gadget-}{\smaller\smaller 2}, after which the produced snapshots are compared. Besides subdirectories containing tests, the \texttt{tests} directory contains a Bash script called \texttt{environment} which handles the setup and teardown of the test environment for all tests.

If a test produces files when run, it should contain a Bash script named \texttt{clean}, which when executed deletes these files. The \texttt{Makefile} will then execute this script every time you call the \texttt{clean\_tests} target.

All tests are run locally, meaning without the use of \textsc{pbs}, even when working remotely. The list of available tests is as follows.
\begin{description}[font=\normalfont\ttfamily]
    \item[basic] \hfill \\
    Performs a basic test of the \concept environment, making sure that all of the software is correctly built and linked. It simply runs the code without any parameter file in both compiled and pure Python mode, using 1, 2 and 4 processes.
    \item[concept\_vs\_gadget\_P3M] \hfill \\
    Performs a comparison test between \concept's $\text{P}^{3}\text{M}$ and \textsc{gadget-}{\smaller\smaller 2}'s TreePM implementation. Only a single process is used.
    \item[concept\_vs\_gadget\_PM] \hfill \\
    Performs a comparison test between \concept's and \textsc{gadget-}{\smaller\smaller 2}'s PM implementations. Since \textsc{gadget-}{\smaller\smaller 2} do not expose its pure PM method to the user, edited versions of the \textsc{gadget-}{\smaller\smaller 2} source files \texttt{pm\_periodic.c} and \texttt{timestep.c} are used. Only a single process is used.
    \item[concept\_vs\_gadget\_PP] \hfill \\
    Performs a comparison test between \concept's PP and \textsc{gadget-}{\smaller\smaller 2}'s tree implementation. Only a single process is used.
    \item[drift] \hfill \\
    Performs a test of \textsc{co\textsl{n}cept}'s drift operation, by comparing it to that of \textsc{gadget-}{\smaller\smaller 2}. ICs containing a particle configuration with no net forces and equal initial velocities are constructed. The particle trajectories are now independent of gravity, making it possible to test the comoving equations of motions. Since gravity is not simply turned off, but rather delicately balanced to produce no net effect, its periodicity is also tested during this test. To test periodicity further, the particles will go trough the side of the box during the simulation. The PP method is used for gravity, and only a single process is used.
    \item[gadget] \hfill \\
    As \textsc{gadget-}{\smaller\smaller 2} is used by many of the other tests, it is important that \textsc{gadget-}{\smaller\smaller 2} itself works properly. This test is simply a shortened version of \textsc{gadget-}{\smaller\smaller 2}'s own \texttt{lcdm\_gas} test.
    \item[kick\_PP\_with\_Ewald] \hfill \\
    Performs a test of \textsc{co\textsl{n}cept}'s PP implementation, by comparing it to \textsc{gadget-}{\smaller\smaller 2}'s tree implementation. ICs containing a particle configuration with net forces only in the $x$-direction are constructed. The particles start out with zero velocities, making the study of the kick operation in isolation easy. Any errors in the implementation of the Ewald method should be detected.
    \item[kick\_PP\_without\_Ewald] \hfill \\
    Performs a test of \textsc{co\textsl{n}cept}'s PP implementation with Ewald summation disabled, by comparing it to \textsc{gadget-}{\smaller\smaller 2}'s tree implementation. \textsc{Gadget-}{\smaller\smaller 2} do not provide a switch to turn off the Ewald summation. To achieve this, the \textsc{gadget-}{\smaller\smaller 2} source file \texttt{forcetree.c} is edited accordingly. ICs containing a particle configuration with net forces only in the $x$-direction are constructed. The particles start out with zero velocities, making the study of the kick operation in isolation easy. Any simple errors in the PP algorithm, having to do with units and the like, should be detected.
    \item[nprocs\_P3M] \hfill \\
    Performs a test of the $\text{P}^{3}\text{M}$ implementation, comparing runs using different numbers of processes. Specifically, 1, 2 4 and 8 processes will be used.
    \item[nprocs\_PM] \hfill \\
    Performs a test of the PM implementation, comparing runs using different numbers of processes. Specifically, 1, 2, 4 and 8 processes will be used.
    \item[nprocs\_PP] \hfill \\
    Performs a test of the PP implementation, comparing runs using different numbers of processes. Specifically, 1, 2, 4 and 8 processes will be used.
    \item[powerspec] \hfill \\
    Performs a test of the power spectrum computation by comparing $\sigma_8$ from this computation with an estimate obtained by taking the rms of the number of particles within cubes the size of spheres with radius \SI{8}{Mpc}.
    \item[pure\_python\_P3M] \hfill \\
    Performs a test of the $\text{P}^{3}\text{M}$ implementation, comparing a compiled run on a single process to pure Python runs with different numbers of processes. Specifically, 1, 2, and 4 processes will be used. Deviations between the compiled run and the pure Python runs signals an error in the translation from Python to C.
    \item[pure\_python\_PM] \hfill \\
    Performs a test of the PM implementation, comparing a compiled run on a single process to pure Python runs with different numbers of processes. Specifically, 1, 2, and 4 processes will be used. Deviations between the compiled run and the pure Python runs signals an error in the translation from Python to C.
    \item[pure\_python\_PP] \hfill \\
    Performs a test of the PP implementation, comparing a compiled run on a single process to pure Python runs with different numbers of processes. Specifically, 1, 2, and 4 processes will be used. Deviations between the compiled run and the pure Python runs signals an error in the translation from Python to C.
    \item[render] \hfil \\
    Generates a random snapshot and renders it first using 1 process and giving the render utility the exact path to the snapshot (with and without specifying a render parameter file). Two copies of this snapshot is then placed in a separate directory. Using 2 processes, the render utility is then given the path to this directory, which should produce a render for each snapshot. Different render parameters are used for the two calls to the render utility. The test will fail if the two renders of the identical snapshots are not themselves identical. The image resolution is also checked. Finally, it is checked whether the text stating the scale factor is clearly visible on both dark and bright backgrounds.
\end{description}
All of the available tests can be run one after another, by
\begin{lstlisting}[language=bash, backgroundcolor=\color{yellow!20}, xleftmargin=\parindent, xrightmargin=\parindent, basicstyle=\small\ttfamily, columns=fullflexible, keepspaces=true, mathescape]
./concept -t all
\end{lstlisting}
This is done as the last step of the \concept installation when using the \texttt{installer}, unless the \texttt{{-}{-}fast} option is given.

\end{document}